\documentclass[prd,preprint]{revtex4}%
\usepackage{amssymb}
\usepackage{amsfonts}
\usepackage{amsmath}
\usepackage{graphicx,color}
\begin{document}
\title{Hairy black hole entropy \\ and the role of solitons in three dimensions}
\author{Francisco Correa$^{1}$, Cristi\'{a}n Mart\'{\i}nez$^{1,2}$, and Ricardo
Troncoso$^{1,2}$}
\email{correa, martinez, troncoso@cecs.cl}
\affiliation{\small $^{1}$Centro de Estudios Cient\'{\i}ficos (CECs), Av.~Arturo Prat 514, Valdivia, Chile \\
$^{2}$Universidad Andr\'{e}s Bello, Av.~Rep\'{u}blica 440, Santiago, Chile}
\preprint{CECS-PHY-11/08}

\begin{abstract}
Scalar fields minimally coupled to General Relativity in three dimensions are
considered. For certain families of self-interaction potentials, new exact solutions
describing solitons and hairy black holes are found. It is shown that they fit
within a relaxed set of asymptotically AdS boundary conditions, whose
asymptotic symmetry group coincides with the one for pure gravity and its
canonical realization possesses the standard central extension. Solitons are
devoid of integration constants and their (negative) mass, fixed and
determined by nontrivial functions of the self-interaction couplings, is shown
to be bounded from below by the mass of AdS spacetime. Remarkably, assuming
that a soliton corresponds to the ground state of the sector of the theory for
which the scalar field is switched on, the semiclassical entropy of the
corresponding hairy black hole is exactly reproduced from Cardy formula once
nonvanishing lowest eigenvalues of the Virasoro operators are taking into
account, being precisely given by the ones associated to the soliton.

This provides further evidence about the robustness of previous results, for
which the ground state energy instead of the central charge appears to play
the leading role in order to reproduce the hairy black hole entropy from a
microscopic counting.
\end{abstract}
\maketitle

\section{Introduction}

The microscopic origin of black hole entropy, a question raised right after
the pioneering work of Bekenstein and Hawking during the 1970's, still remains
as a challenging unsolved puzzle. Nevertheless, the most compelling current
proposals appear to converge in the sense that, regardless the precise
mechanisms and assumptions, an emergent conformal symmetry in two dimensions,
endowed with a suitable central extension, allows to reproduce the
semiclassical entropy of different classes of black holes from a microscopic
counting (see e.g. \cite{Strominger-Vafa, Strominger, BSS, Carlip-conf,
Kerr-CFT}, as well as \cite{Carlip-last, Cvetic-Larsen1, Cvetic-Larsen2} and
references therein).

One of the simplest and clear examples is the one provided by Strominger
\cite{Strominger}. This proposal relies on an observation pushed forward by
Brown and Henneaux during the 1980's \cite{Brown-Henneaux} and currently
interpreted in terms of the AdS/CFT correspondence
\cite{Maldacena-Klebanov-Witten}. As follows from \cite{Brown-Henneaux}, since
the asymptotic symmetries of General Relativity with negative cosmological
constant in three dimensions correspond to two copies of the Virasoro algebra,
a consistent quantum theory of gravity should then be described in terms of a
conformal field theory in two dimensions, with a central charge given by%
\begin{equation}
c=\frac{3l}{2G}\ , \label{central charge}%
\end{equation}
where $G$ and $l$ stand for the Newton constant and the AdS radius,
respectively. Thus, in \cite{Strominger} it was assumed that if the CFT
fulfills some physically sensible properties, the physical states form a
consistent unitary representation of the conformal algebra, so that the
asymptotic growth of the number of states must be given by Cardy formula
\cite{Cardy}. Remarkably, the result precisely agrees with semiclassical
entropy of the BTZ black hole \cite{BTZ, BHTZ} provided the central charge is
exactly given by (\ref{central charge}).

Nonetheless, there are known examples for which this proposal has to be
refined, since for them the central charge does not play the leading role in
order to reproduce the semiclassical black hole entropy from a microscopic
counting. Indeed, as explained in \cite{CMT} the asymptotic growth of the
number of states can be expressed only in terms of the spectrum of the
Virasoro operators without making any explicit reference to the central
charges, so that the relevant quantities that allow to reproduce the black
hole entropy turn out to be the lowest eigenvalues of the Virasoro operators.
This can be seen as follows. If the spectrum of the Virasoro operators
$L_{0}^{\pm}$, whose eigenvalues are given by $\Delta^{\pm}$, is such that
their lowest eigenvalues, denoted by $\Delta_{0}^{\pm}$, are nonvanishing
(i.e., for $\Delta_{0}^{\pm}\neq0$), Cardy formula reads (see e.g.
\cite{Cardy, Carlip, MuInPark, iran})%
\begin{equation}
S=2\pi\sqrt{\frac{\left(  c^{+}-24\Delta_{0}^{+}\right)  }{6}\left(
\Delta^{+}-\frac{c^{+}}{24}\right)  }+2\pi\sqrt{\frac{\left(  c^{-}%
-24\Delta_{0}^{-}\right)  }{6}\left(  \Delta^{-}-\frac{c^{-}}{24}\right)  }\ ,
\label{Cardy central reloaded}%
\end{equation}
where it is assumed that the ground state is non degenerate. As pointed out in
\cite{CMT}, \emph{on a cylinder}, the zero mode of the Virasoro operators gets
shifted according to%
\begin{equation}
\tilde{L}_{0}^{\pm}:=L_{0}^{\pm}-\frac{c^{\pm}}{24}\ ,
\label{Shifted Virasoro}%
\end{equation}
so that formula (\ref{Cardy central reloaded}) can be naturally written as%
\begin{equation}
S=4\pi\sqrt{-\tilde{\Delta}_{0}^{+}\tilde{\Delta}^{+}}+4\pi\sqrt
{-\tilde{\Delta}_{0}^{-}\tilde{\Delta}^{-}}\ , \label{Cardy super reloaded}%
\end{equation}
where $\tilde{\Delta}^{\pm}$ correspond to the eigenvalues of $\tilde{L}%
_{0}^{\pm}$, while $\tilde{\Delta}_{0}^{\pm}$ stand for the lowest ones.
Therefore, from (\ref{Cardy super reloaded}) it is apparent that the
asymptotic growth of the number of states can be precisely obtained if one
only knew the spectrum of $\tilde{L}_{0}^{\pm}$ without knowledge about the
central charges.

Note that when the lowest eigenvalues of the Virasoro operator vanish, i.e.
for $\Delta_{0}^{\pm}=0$, or equivalently $\tilde{\Delta}_{0}^{\pm}%
=-\frac{c^{\pm}}{24}$, formula (\ref{Cardy super reloaded}) reduces to its
more familiar form, given by,
\begin{equation}
S=2\pi\sqrt{\frac{c^{+}}{6}\tilde{\Delta}^{+}}+2\pi\sqrt{\frac{c^{-}}{6}%
\tilde{\Delta}^{-}}\ . \label{Cardy simple}%
\end{equation}
In this case, as shown in \cite{Strominger}, assuming that the eigenvalues of
$\tilde{L}_{0}^{\pm}$ are given by the corresponding canonical generators
according to%
\begin{equation}
\tilde{\Delta}^{\pm}=\frac{1}{2}(Ml\pm J)\ , \label{delta mas menos}%
\end{equation}
where $M$ and $J$ stand for the mass and the angular momentum, respectively,
then formula (\ref{Cardy simple}) precisely reproduces the semiclassical
entropy of the BTZ black hole.

Therefore, for instances such that the lowest eigenvalues of the Virasoro
operators do not vanish, i.e., for $\Delta_{0}^{\pm}\neq0$, formula
(\ref{Cardy simple}) does not apply. Remarkably, as explained in \cite{CMT},
in these cases the semiclassical black hole entropy can still be successfully
reproduced by virtue of the generic formula given by
(\ref{Cardy super reloaded}) once the ground state configuration is suitably
identified. A concrete example where this effect occurs is provided by the
existence of hairy black holes found in \cite{HMTZ-2+1} for General Relativity
minimally coupled to a self-interacting scalar field in three dimensions. The
action is given by%
\begin{equation}
I[g_{\mu\nu},\phi]=\frac{1}{\pi G}\int d^{3}x\sqrt{-g}\left[  \frac{R}%
{16}-\frac{1}{2}(\nabla\phi)^{2}-V(\phi)\right]  \;, \label{Action}%
\end{equation}
and the self-interaction potential, expanded around $\phi=0$, is assumed to be
of the form
\begin{equation}
V(\phi)={-\frac{1}{8l^{2}}-\frac{3}{8l^{2}}\phi^{2}-\frac{1}{2l^{2}}\phi^{4}%
}+\mathcal{O}(\phi^{6})\ , \label{Generic Potential}%
\end{equation}
so that the first term corresponds to the cosmological constant $\Lambda
=-1/l^{2}$, while the second one is the mass term, with $m^{2}=-3/(4l^{2})$,
fulfilling the Breitenlohner-Freedman bound \cite{BF,MT}. As shown in
\cite{HMTZ-2+1}, the scalar field is able to acquire slow fall-off at
infinity, so that the action (\ref{Action}) admits a suitable set of
asymptotically AdS boundary conditions in a relaxed sense as compared with the
one of Brown and Henneaux \cite{Brown-Henneaux}, which nevertheless possesses
the same asymptotic symmetries, i.e., they are also left invariant under the
conformal group in two dimensions, spanned by two copies of the Virasoro
algebra. The asymptotic conditions are given by:%
\begin{equation}
\phi=\frac{\chi}{r^{1/2}}+\alpha\frac{\chi^{3}}{r^{3/2}}+O(r^{-5/2})
\label{asympt scalar general}%
\end{equation}%
\begin{equation}%
\begin{array}
[c]{lll}%
g_{rr}=\displaystyle\frac{l^{2}}{r^{2}}-\frac{4l^{2}\chi^{2}}{r^{3}}%
+O(r^{-4}) &  & \displaystyle g_{tt}=-\frac{r^{2}}{l^{2}}+O(1)\\[2mm]%
g_{tr}=O(r^{-2}) &  & g_{\varphi\varphi}=r^{2}+O(1)\\[1mm]%
g_{\varphi r}=O(r^{-2}) &  & g_{t\varphi}=O(1)
\end{array}
\label{asympt metric general}%
\end{equation}
where $\chi=\chi(t,\varphi)$ is an arbitrary function, and $\alpha$ is an
arbitrary constant. One of the effects of relaxing the asymptotic conditions
is reflected through the fact that the generators of the asymptotic symmetries
acquire a nontrivial contribution from the scalar field. Following the
Regge-Teitelboim approach \cite{Regge-Teitelboim}, the canonical generators
were found to be given by%
\begin{equation}
Q(\xi)=\frac{1}{16\pi G}\!\int\!\!d\varphi\left\{  \frac{\xi^{\bot}}%
{lr}\!\left(  (g_{\varphi\varphi}-r^{2})\!-\!2r^{2}(lg^{-1/2}-1)\right)
\!+\!2\xi^{\varphi}\pi_{\ \varphi}^{r}\!+\!\xi^{\bot}\frac{2r}{l}\!\left[
\phi^{2}-2l\frac{\phi\partial_{r}\phi}{\sqrt{g_{rr}}}\right]  \right\}  ,
\label{Q general}%
\end{equation}
where the reference background is chosen to be the massless BTZ black hole.
The corresponding Poisson brackets were shown to span two copies of the
Virasoro algebra with the standard central charges $c^{+}=c^{-}=c$, where $c$
is given by eq. (\ref{central charge}).

An exact analytic hairy black hole solution whose asymptotic behavior fits
within the boundary conditions given by eqs. (\ref{asympt scalar general}),
(\ref{asympt metric general}) was found in \cite{HMTZ-2+1} for the following
self-interaction potential%
\begin{equation}
V_{1,\nu}(\phi)=-\frac{1}{8l^{2}}\left(  \cosh^{6}\phi+\nu\sinh^{6}%
\phi\right)  \;, \label{Potential}%
\end{equation}
which belongs to the class defined in (\ref{Generic Potential}). As shown in
\cite{CMT}, for this specific potential, the field equations corresponding to
(\ref{Action}) also admit an exact analytic soliton solution, being such that
the metric and the scalar field are regular everywhere and fulfill the
boundary conditions (\ref{asympt scalar general}) and
(\ref{asympt metric general}). The soliton turns out to be devoid of
integration constants, and it has a precise fixed (negative) mass $M_{0}$
determined by the Newton constant and the self-interaction parameter $\nu$.
This fact naturally suggests to regard the soliton as the ground state of the
\textquotedblleft hairy sector\textquotedblright, for which the scalar field
is switched on. Remarkably, assuming that the lowest eigenvalues of the
Virasoro operators are determined by the global charges of the soliton,
according to eq. (\ref{delta mas menos}), i.e., $\tilde{\Delta}_{0}^{\pm
}=\frac{l}{2}M_{0}$, the asymptotic growth of the number of states, given by
eq. (\ref{Cardy super reloaded}), reduces to%
\begin{equation}
S=4\pi l\sqrt{-M_{0}M}\ , \label{Cardy reloaded static}%
\end{equation}
where $M=\frac{2}{l}\tilde{\Delta}^{\pm}$ corresponds to the mass of
the hairy black hole, which exactly reproduces its
semiclassical entropy $S=\frac{A}{4G}$. 

One may wonder whether this result is just a curiosity of the particular model
specified by the potential in (\ref{Potential}), or actually corresponds to a
generic feature of hairy black holes. In this article we construct new examples
that strongly support the latter possibility. This is carried out for
different self-interaction potentials within the class in eq.
(\ref{Generic Potential}), being simultaneously involved enough so as to
provide non trivial lowest eigenvalues for the Virasoro operators in the hairy
sector, as well as sufficiently simple in order to find exact analytic hairy
black holes and their corresponding solitons.

In what follows we show the existence of new analytic hairy black hole and
soliton solutions for different classes of self-interaction potentials. In the
next section a one-parameter family of potentials which differs from
(\ref{Potential}) is considered, while in section \ref{Section3} a class of
potentials that depend on two parameters is discussed. Section \ref{Section4}
is devoted to the explicit microscopic computation of the entropy of the hairy
black holes mentioned above in terms of their corresponding solitons by means
of formula (\ref{Cardy super reloaded}) which reduces to eq.
(\ref{Cardy reloaded static}) in the static case.  Final remarks are given 
in section \ref{discussion}. Appendix \ref{apendiceA} includes a description
of certain analytic functions that become relevant in order to describe the
properties of hairy black holes and solitons, and finally, appendix B is
devoted to discuss the new exact solutions in the conformal (Jordan) frame.

\section{Case 1: Hairy black holes and solitons for a uniparametric family of
potentials}

\label{Section2}

Let us consider the following class of self-interaction potentials,
\begin{equation}
V_{0,\nu}(\phi)=-\frac{1}{16l^{2}}(\nu\cosh^{8}\phi-\nu\cosh^{4}\phi
+2\cosh^{6}\phi\left(  1-\nu\ln\cosh^{2}\phi\right)  )\,,\label{V 0 nu}%
\end{equation}
which, apart from the AdS radius $l$, depends on a single parameter $\nu$.
Around $\phi=0$, the potential behaves as
\begin{equation}
V_{0,\nu}(\phi)={-\frac{1}{8l^{2}}-\frac{3}{8l^{2}}\phi^{2}-\frac{1}{2l^{2}%
}\phi^{4}-\frac{94+5\nu}{240l^{2}}\phi^{6}}+\mathcal{O}(\phi^{8}%
)\ ,\label{asympo1}%
\end{equation}
and it then falls within the family defined in eq. (\ref{Generic Potential})
that is consistent with the boundary conditions given by
(\ref{asympt scalar general}) and (\ref{asympt metric general}). Note that
$V_{0,\nu}(\phi)$ does not overlap with $V_{1,\nu}(\phi)$ in eq.
(\ref{Potential}) for any value of $\nu$. As mentioned above, the self
interaction (\ref{V 0 nu}) turns out to be involved enough so as to possess a
ground state in the hairy sector with nontrivial lowest eigenvalues for the
Virasoro operators, but nevertheless it becomes simple in order to produce
exact analytic hairy black holes and solitons. This is discussed next.

\subsection{Black hole}

\label{Black hole I}

The field equations that correspond to the action (\ref{Action}) with
$V=V_{0,\nu}(\phi)$ admit an exact solution, whose the line element reads
\begin{equation}
ds^{2}=-\frac{r^{2}}{l^{2}}h(r)dt^{2}+\frac{l^{2}r^{2}dr^{2}}{(r+a)^{4}%
h(r)}+r^{2}d\varphi^{2}\,, \label{einstein}%
\end{equation}
with
\begin{equation}
h(r)=1+\nu\left(  \frac{a}{r+a}+\ln\frac{r}{r+a}\right)  \ ,
\label{h enchulado}%
\end{equation}
and the scalar field is given by
\begin{equation}
\phi=\mathrm{arctanh}\,\sqrt{\frac{a}{r+a}}\,. \label{einstein scalar}%
\end{equation}
The coordinates range as $-\infty<t<\infty$, $r>0$, $0\leq\varphi<2\pi$, and
the solution depends on a single non-negative integration constant $a$.
Remarkably, the scalar field is regular everywhere and the solution describes
a hairy black hole provided $\nu>0$, otherwise the singularity at the origin
$r=0$ becomes naked. There is a single horizon located at
\begin{equation}
r_{+}=a\Phi_{\nu}\ ,
\end{equation}
where $\Phi_{\nu}$ depends only on the parameter $\nu$ and it is given by%
\begin{equation}
\Phi_{\nu}:=\frac{-W(-e^{-1-\frac{1}{\nu}})}{1+W(-e^{-1-\frac{1}{\nu}})}\ .
\label{phinu}%
\end{equation}
Here $W$ stands for the Lambert $W$ function, defined as
\begin{equation}
W(z)e^{W(z)}=z\ ,
\end{equation}
which for $-1/e<z<0$ has an upper branch that ranges according to $-1<W(z)<0$
(see e.g. \cite{libro})\footnote{Note that, for $-1/e<z<0$, the lower branch
of the Lambert $W$ function ranges as $-\infty<W(z)<-1$. Hereafter the lower
branch is not considered since the solution would describe a naked singularity
and it would then be ruled out by cosmic censorship.}. For latter purposes it
is worth pointing out that for $0<\nu<\infty$, the function $\Phi_{\nu}$ is
bounded as%
\begin{equation}
0<\Phi_{\nu}^{2}<\frac{\nu}{2}\ . \label{Bound Phi}%
\end{equation}
Further details about $\Phi_{\nu}$ are included in appendix \ref{apendiceA}.


The Hawking temperature of the hairy black hole (\ref{einstein}),
(\ref{einstein scalar}) can be obtained demanding regularity of the Euclidean
solution at the horizon and it is found to be proportional to the integration
constant $a$, which reads
\begin{equation}
T=\frac{a\nu}{4\pi l^{2}\Phi_{\nu}}\ ,
\end{equation}
while its entropy is given by%
\begin{equation}
S=\frac{A}{4G}=\frac{\pi r_{+}}{2G}=\frac{\pi\Phi_{\nu}}{2G}a\ .
\label{entropy uni}%
\end{equation}
Asymptotically, this hairy black hole behaves as%
\begin{equation}
\phi=\frac{a^{1/2}}{r^{1/2}}-\frac{1}{6}\frac{a^{3/2}}{r^{3/2}}+\mathcal{O}%
\left(  \frac{1}{r^{5/2}}\right)  \ , \label{asym1sca}%
\end{equation}
with
\begin{align}
g_{tt}  &  =-\frac{r^{2}}{l^{2}}+\frac{a^{2}\nu}{2l^{2}}+\mathcal{O}\left(
\frac{1}{r}\right)  \ ,\label{asym1me}\\
g_{rr}  &  =\frac{l^{2}}{r^{2}}-\frac{4l^{2}a}{r^{3}}+\mathcal{O}\left(
\frac{1}{r^{4}}\right)  \ ,\\
g_{\varphi\varphi}  &  =r^{2}\ ,
\end{align}
and then falls within the set defined by eqs. (\ref{asympt scalar general}),
(\ref{asympt metric general}) with $\chi=\sqrt{a}$ and $\alpha=-1/6$.
Therefore, the solution is asymptotically AdS in a relaxed sense as compared
with the standard one \cite{Brown-Henneaux}. The mass of the hairy black hole
can be readily computed by virtue of eq. (\ref{Q general}),
yielding\footnote{It would be interesting to compare this result with the ones
that could be obtained from different approaches that are adapted to deal with
scalar fields and relaxed AdS asymptotics, as the ones in Refs. \cite{Glenn1,
Glenn2, GMT}. Further results along the lines of \cite{Clement, Turkey,
Banados-Theisen, Myung, Lashkari}, previously found for the hairy black hole
of Ref. \cite{HMTZ-2+1}, would also worth to be explored for the new solutions
found here.}%
\begin{equation}
M=Q(\partial_{t})=\frac{\nu a^{2}}{16Gl^{2}}\ . \label{M1}%
\end{equation}

Note that the scalar field cannot be switched off keeping the mass fixed.
Indeed, the solution depends on a single integration constant, so that for
$\phi \to 0$, the geometry approaches the one of the massless BTZ
black hole.

\subsection{Soliton}

\label{soliton1}

For the self-interaction potential $V_{0,\nu}(\phi)$ in eq. (\ref{V 0 nu}),
the field equations also admit the following solution:
\begin{equation}
ds^{2}=-\frac{r^{2}}{l^{2}}dt^{2}+\frac{l^{2}r^{2}dr^{2}}{(r+\frac{2l\Phi
_{\nu}}{\nu})^{4}H(r)}+r^{2}H(r)d\varphi^{2}\,, \label{Soliton I}%
\end{equation}
with
\begin{equation}
H(r)=1+\frac{2l\Phi_{\nu}}{r+\frac{2l\Phi_{\nu}}{\nu}}+\nu\ln\frac{r}%
{r+\frac{2l\Phi_{\nu}}{\nu}}\ , \label{H(r)}%
\end{equation}
and%
\begin{equation}
\phi(r)=\mathrm{arctanh}\,\sqrt{\frac{1}{1+\displaystyle\frac{\nu r}%
{2l\Phi_{\nu}}}}\ , \label{Phi I}%
\end{equation}
where the coordinates range according to $-\infty<t<\infty$, $0\leq
\varphi<2\pi$, and%
\begin{equation}
\frac{2l\Phi_{\nu}^{2}}{\nu}\leq r<\infty\ ,
\end{equation}
with $\nu>0$. Note that the solution is devoid of integration constants and
depends only on the self-interaction parameter $\nu$ and the AdS radius $l$.
It is simple to verify that the solution is smooth and regular everywhere.
Indeed, the behaviour of the solution around the origin, located at
$r=\frac{2l\Phi_{\nu}^{2}}{\nu}$ can be seen from the expansion of
(\ref{H(r)}), given by%
\[
H\left(  r\right)  =\frac{\nu^{2}}{2l\Phi_{\nu}^{2}(1+\Phi_{\nu})^{2}}\left(
r-\displaystyle\frac{2l\Phi_{\nu}^{2}}{\nu}\right)  +\mathcal{O}\left[
\left(  r-\displaystyle\frac{2l\Phi_{\nu}^{2}}{\nu}\right)  ^{2}\right]  \ ,
\]
so that defining $\hat{t}=\frac{2\Phi_{\nu}^{2}}{\nu}t$, and $\rho^{2}%
=r^{2}H(r)$, the metric approaches to the one of Minkowski spacetime,%

\[
ds^{2}\rightarrow-d\hat{t}^{2}+d\rho^{2}+\rho^{2}d\varphi^{2}\ .
\]
Analogously, the form of the scalar field around $\rho=0$ is given by%
\[
\phi(\rho)=\mathrm{arctanh}\,\sqrt{\frac{1}{1+\Phi_{\nu}}}-\frac{\nu\left(
1+\Phi_{\nu}\right)  ^{3/2}}{8l^{2}\Phi_{\nu}^{4}}\rho^{2}+\mathcal{O}\left(
\mathcal{\rho}^{4}\right)  \ .
\]

The asymptotic behavior of (\ref{Phi I}) and (\ref{Soliton I}) is given by%
\begin{equation}
\phi=\left(  \frac{2l\Phi_{\nu}}{\nu}\right)  ^{1/2}\frac{1}{r^{1/2}}-\frac
{1}{6}\left(  \frac{2l\Phi_{\nu}}{\nu}\right)  ^{3/2}\frac{1}{r^{3/2}%
}+\mathcal{O}\left(  \frac{1}{r^{5/2}}\right)  \ ,
\end{equation}
with
\begin{align}
g_{tt}  &  =-\frac{r^{2}}{l^{2}}\ ,\nonumber\\
g_{rr}  &  =\frac{l^{2}}{r^{2}}-\frac{2l\Phi_{\nu}}{\nu}\frac{4l^{2}}{r^{3}%
}+\mathcal{O}\left(  \frac{1}{r^{4}}\right)  \ ,\\
g_{\varphi\varphi}  &  =r^{2}+\mathcal{O}(1)\ ,\nonumber
\end{align}
and then belongs to the class of relaxed asymptotically AdS conditions defined
by eqs. (\ref{asympt scalar general}) and (\ref{asympt metric general}) with
$\chi=\left(  \frac{2l\Phi_{\nu}}{\nu}\right)  ^{1/2}$ and, as for the hairy
black hole discussed in \ref{Black hole I}, $\alpha=-1/6$. The mass of this
solution can then be obtained by virtue of eq. (\ref{Q general}) which yields%
\begin{equation}
M_{\mathrm{sol}}=-\frac{\Phi_{\nu}^{2}}{4G\nu}\ . \label{Msol I}%
\end{equation}
In sum, this solution is regular everywhere, shares the same causal structure
with AdS spacetime, and since it has a fixed finite mass, it describes a soliton.

Note that by virtue of (\ref{Bound Phi}), which holds for the allowed range of
the self-interaction coupling, $\nu>0$, the soliton mass $M_{0}%
=M_{\mathrm{sol}}$ becomes bounded according to%
\begin{equation}
-\frac{1}{8G}<M_{0}<0\ .\label{boundgroundstate}%
\end{equation}

\section{Case 2: Hairy black holes and solitons for self-interaction
potentials with two parameters.}

\label{Section3}

Here we consider a wider class of self-interaction potentials being such that
not only interpolates, but generalizes the ones considered above. This is
given by%
\begin{align}
V_{\lambda,\nu}(\phi)  &  =-\frac{\nu}{16l^{2}}\sinh^{2}\phi\left[
\lambda(\lambda+1)-2\lambda(\lambda+2)\cosh^{2}\phi+(1+4\lambda+\lambda
^{2})\cosh^{4}\phi\right.  \left.  -(\lambda-1)\cosh^{6}\phi\right]
\nonumber\label{V lambda nu}\\
&  -\frac{\cosh^{6}\phi-\lambda^{2}\sinh^{6}\phi}{8l^{2}(\lambda-1)}\left[
(\lambda-1)+\nu\ln(\lambda-(\lambda-1)\cosh^{2}\phi)\right]  ,
\end{align}
which depends on two parameters $\nu$, $\lambda$, and $l$ stands for the AdS
radius. The behavior of $V_{\lambda,\nu}$ around $\phi=0$ reads%
\begin{equation}
V_{\lambda,\nu}(\phi)\xrightarrow[ \phi \rightarrow{}0]\,{-\frac{1}{8l^{2}%
}-\frac{3}{8l^{2}}\phi^{2}-\frac{1}{2l^{2}}\phi^{4}-\frac{94-30\lambda
^{2}+5\lambda^{2}\nu-10\nu\lambda+5\nu}{240l^{2}}\phi^{6}}+\mathcal{O}%
(\phi^{8})\,,
\end{equation}
so that it belongs to the class defined in eq. (\ref{Generic Potential}). Note
that for $\lambda=0$ the self interaction (\ref{V lambda nu}) reduces to
$V_{0,\nu}(\phi)$ in Eq. (\ref{V 0 nu}), while after redefining $\nu
=6\frac{\tilde{\nu}+1}{(\lambda-1)^{2}}$, in the limit $\lambda\rightarrow1$,
the potential (\ref{V lambda nu}) acquires the form of $V_{1,\tilde{\nu}}%
(\phi)$ in eq. (\ref{Potential}).

In what follows, exact hairy black holes and soliton solutions for this self
interaction are explicitly found.

\subsection{Black hole}

\label{Black hole II}

In the case of $V=V_{\lambda,\nu}(\phi)$, the field equations possess an
analytic solution. The metric is given by
\begin{equation}
ds^{2}=\Omega^{2}(r)\left(  -f(r)dt^{2}+\frac{dr^{2}}{f(r)}+r^{2}d\varphi
^{2}\right)  \,, \label{Twobh}%
\end{equation}
where
\begin{align}
\Omega^{2}(r)  &  =\left(  \frac{\lambda(r+b)-b}{\lambda(r+b)}\right)
^{2}\ ,\nonumber\\
f(r)  &  =\frac{r^{2}}{l^{2}}-\frac{\nu}{\lambda(\lambda-1)l^{2}}\left(
\frac{(\lambda-1)^{2}}{2}b^{2}-b(\lambda-1)r+\lambda r^{2}\ln\left(
1+b\frac{\lambda-1}{\lambda r}\right)  \right) \,,  \label{fr}%
\end{align}
and the scalar field reads,
\begin{equation}
\phi=\mathrm{arctanh}\,\sqrt{\frac{b}{\lambda(r+b)}}\,. \label{Twosc}%
\end{equation}
The solution depends on a single integration constant $b$, and the coordinates
range as $-\infty<t<\infty$, $0\leq\varphi<2\pi$, and $r>r_{s}$, where
$r=r_{s}$ stands for the location of the curvature singularity specified
below. The asymptotic conditions (\ref{asympt scalar general}%
),\ (\ref{asympt metric general}) are also fulfilled with $\alpha=-\frac{1}%
{6}(1+3\lambda)$ and $\chi=(b/\lambda)^{1/2}$, which means that the hairy
solution is well defined provided $b/\lambda>0$. Indeed, making the shift
$r\rightarrow r+\frac{b}{\lambda}$,  the asymptotic behavior of the
solution reads
\begin{equation}
\phi=\frac{(b/\lambda)^{1/2}}{r^{1/2}}-\frac{1}{6}(1+3\lambda)\frac
{(b/\lambda)^{3/2}}{r^{3/2}}+\mathcal{O}\left(  \frac{1}{r^{5/2}}\right)
\label{asym2sca}%
\end{equation}%
\[%
\begin{array}
[c]{lll}%
g_{rr}=\displaystyle\frac{l^{2}}{r^{2}}-\frac{b}{\lambda}\frac{4l^{2}}{r^{3}%
}+O(r^{-4}) &  & \displaystyle g_{tt}=-\frac{r^{2}}{l^{2}}+O(1)\\[2mm]%
g_{tr}=O(r^{-2}) &  & g_{\varphi\varphi}=r^{2}+O(1)\\[1mm]%
g_{\varphi r}=O(r^{-2}) &  & g_{t\varphi}=O(1)
\end{array}
\]
The mass can then be computed by virtue of eq. (\ref{Q general}), which gives
\begin{equation}
M=\frac{b^{2}}{16Gl^{2}}\frac{(\lambda-1)^{2}}{\lambda^{2}}\nu\ . \label{mii}%
\end{equation}

As in the case of the solution found in the previous section, this one
depends on a single integration constant $b$, and it is such that the
massless BTZ black hole in vacuum ($\phi=0$) is recovered for $b \to 0$.

It is simple to verify that the solution describes a hairy black hole with a
regular scalar field on and outside the event horizon provided $\nu>0$, which
ensures the mass (\ref{mii}) is positive. Thus, remarkably, although the self
interaction looks somehow involved, positivity of the hairy
black hole energy  still goes by hand with cosmic censorship.

The event horizon is located at $r=r_{+}$, with
\begin{equation}
\label{horizon2}r_{+}=\frac{b}{\lambda}(1-\lambda)\mathcal{R}_{\lambda,\nu}\ ,
\end{equation}
where the function $\mathcal{R}_{\lambda,\nu}$ is defined as the real root of
\begin{equation}
\label{rdef}\mathcal{R}_{\lambda,\nu}^{2}-\frac{\nu}{(\lambda-1)}\left[
\frac{\lambda}{2}+\mathcal{R}_{\lambda,\nu}+\mathcal{R}_{\lambda,\nu}^{2}%
\log\left(  1-\frac{1}{\mathcal{R}_{\lambda,\nu}}\right)  \right]  =0\ ,
\end{equation}
which only holds for $\nu>0$.

In the case of $\lambda>1$ there is a curvature singularity at $r_{s}=0$, and
since the function $\mathcal{R}_{\lambda,\nu}$ ranges as $-\infty
<\mathcal{R}_{\lambda,\nu}<0$, it is always surrounded by the event horizon.

For $\lambda<1$ the curvature singularity is located at $r_{s}=b\frac
{1-\lambda}{\lambda}\,$, and it is also always cloaked by the event horizon
since in this case the function $\mathcal{R}_{\lambda,\nu}$ ranges according
to $1<\mathcal{R}_{\lambda,\nu}<\infty$.

The Hawking temperature and hairy black hole entropy are given by%
\begin{align}
T  &  =\frac{\nu}{4\pi l^{2}}\frac{b}{\lambda}\frac{1-\lambda}{\Upsilon
_{\lambda,\nu}}\ ,\\
S  &  =\frac{A}{4G}=\frac{\pi}{2G}(1-\lambda)\frac{b}{\lambda}\Upsilon
_{\lambda,\nu}\ , \label{Stwo}%
\end{align}
respectively, where the function
\begin{equation}
\Upsilon_{\lambda,\nu}:=\frac{(1-\lambda)\mathcal{R}_{\lambda,\nu}%
(\mathcal{R}_{\lambda,\nu}-1)}{\lambda+(1-\lambda)\mathcal{R}_{\lambda,\nu}%
}\ , \label{Upsilon}%
\end{equation}
fulfills $\Upsilon_{\lambda,\nu}(1-\lambda)>0$, so that the temperature and
the entropy are manifestly positive, and it is bounded as%
\begin{equation}
\Upsilon_{\lambda,\nu}^{2}<\frac{\nu}{2}\ . \label{bound upsilon}%
\end{equation}
Further details about the functions $\mathcal{R}_{\lambda,\nu}$ and
$\Upsilon_{\lambda,\nu}$ are revisited in Appendix \ref{apendiceA}.

\subsection{Soliton}

The field equations for the self-interaction potential $V_{\lambda,\nu}$ in
(\ref{V lambda nu}), with $\nu>0$, also admit the following soliton solution:
\begin{equation}
\phi(r)=\mathrm{arctanh}\,\sqrt{\frac{\gamma_{\lambda,\nu}}{r+(1+\lambda
)\gamma_{\lambda,\nu}}}\,,\label{Phi II}%
\end{equation}
with
\begin{equation}
ds^{2}=-\frac{(r+\gamma_{\lambda,\nu})^{2}(r+\lambda\gamma_{\lambda,\nu})^{2}%
}{l^{2}(r+(1+\lambda)\gamma_{\lambda,\nu})^{2}}dt^{2}+\left(  \frac
{r+\lambda\gamma_{\lambda,\nu}}{r+(1+\lambda)\gamma_{\lambda,\nu}}\right)
^{2}\left(  \frac{dr^{2}}{g(r)}+l^{2}g(r)d\varphi^{2}\right)
\ ,\label{Soliton II}%
\end{equation}
where
\begin{equation}
g(r)=\frac{\nu\,\gamma_{\lambda,\nu}}{l^{2}}\left(  r+(2+\lambda-\lambda
^{2})\frac{\gamma_{\lambda,\nu}}{2}\right)  +\frac{1}{l^{2}}(r+\gamma
_{\lambda,\nu})^{2}\left(  1+\frac{\nu}{1-\lambda}\ln\left(  \frac
{r+\lambda\gamma_{\lambda,\nu}}{r+\gamma_{\lambda,\nu}}\right)  \right)  \ ,
\end{equation}
and $\gamma_{\lambda,\nu}:=\frac{2l\Upsilon_{\lambda,\nu}}{(1-\lambda)\nu}$ is
a two-parametric constant.

The coordinates range according to $-\infty<t<\infty$, $0\leq\varphi<2\pi$,
and
\begin{equation}
\frac{2l\Upsilon_{\nu,\lambda}^{2}}{\nu}\leq r<\infty\ .
\end{equation}
This solution depends only on the parameters of the potential, $\nu$,
$\lambda$ and the AdS radius $l$. Thus, as in the case discussed in Sec.
\ref{soliton1}, the soliton has no integration constants and it is simple to
verify that the solution is smooth and regular everywhere. The soliton also
fulfills the asymptotic conditions in eqs. (\ref{asympt scalar general}) and
(\ref{asympt metric general}), with $\chi=\left(  \gamma_{\lambda,\nu}\right)
^{1/2}$ and $\alpha=-\frac{1}{6}(1+3\lambda)$. Indeed, for $r\rightarrow
\infty$ the solution behaves as%

\begin{equation}
\phi=\frac{\left(  \gamma_{\lambda,\nu} \right)  ^{1/2}}{r^{1/2}}-\frac{1}%
{6}(1+3\lambda)\frac{\left(  \gamma_{\lambda,\nu} \right)  ^{3/2}}{r^{3/2}%
}+\mathcal{O}\left(  \frac{1}{r^{5/2}}\right) \nonumber
\end{equation}

\[%
\begin{array}
[c]{lll}%
g_{rr}=\displaystyle\frac{l^{2}}{r^{2}}- \gamma_{\lambda,\nu} \frac{4l^{2}%
}{r^{3}}+O(r^{-4}) &  & \displaystyle g_{tt}=-\frac{r^{2}}{l^{2}}+O(1)\\[2mm]%
g_{tr}=O(r^{-2}) &  & g_{\varphi\varphi}=r^{2}+O(1)\\[1mm]%
g_{\varphi r}=O(r^{-2}) &  & g_{t\varphi}=O(1)\label{asympt metric copy(1)}%
\end{array}
\]
Note that the constant $\alpha$ coincides with the one of the hairy black hole
for the same potential $V_{\lambda,\nu}$.

The soliton mass can then be readily obtained from eq. (\ref{Q general}),
which gives
\begin{equation}
M_{\mathrm{sol}}=-\frac{\Upsilon_{\nu,\lambda}^{2}}{4G\nu}\ ,
\label{mass lambda nu}%
\end{equation}
and by virtue of (\ref{bound upsilon}) turns out to be bounded exactly as in
eq. (\ref{boundgroundstate}), i.e.,%
\[
-\frac{1}{8G}<M_{0}<0\ .
\]

As a concluding remark of this section, one can verify that in the limits
$\lambda\rightarrow0$ and $\lambda\rightarrow1$, not only the potentials
$V_{0,\nu}$, and $V_{1,\nu}$ in eqs. (\ref{V 0 nu}) and (\ref{Potential}) are
recovered from $V_{\lambda,\nu}$, respectively, but also their corresponding
hairy black holes and solitons described above, as well as those in Refs.
\cite{HMTZ-2+1,CMT}. In the case $\lambda\rightarrow0$, this can be explicitly
see as follows\footnote{In the limit $\lambda\rightarrow1$, following a
similar procedure, the hairy black hole and soliton solutions of
\cite{HMTZ-2+1,CMT} can be recovered from those in Sec. \ref{Section2}.}:

Taking the limit $\lambda\rightarrow0$ with $\frac{b}{\lambda}\rightarrow a$,
the solution defined by the metric (\ref{Twobh}) and the scalar field
(\ref{Twosc}) becomes the black hole solution of section \ref{Black hole I}
given by (\ref{einstein}) and (\ref{einstein scalar}), provided the radial
coordinates is shifted as $r\rightarrow r+a$.

In the case of the soliton solution, the function $\Phi_{\nu}$ is exactly
recovered from the limit $\lambda\rightarrow0$ in (\ref{rdef}) and
(\ref{Upsilon}), i.e. $\Upsilon_{0,\nu}=\Phi_{\nu}$. In this way, the soliton
that corresponds to the two parametric potential case, described by eqs.
(\ref{Soliton II}) and (\ref{Phi II}), reduces to the uniparametric one in
(\ref{Soliton I}) and (\ref{Phi I}).

In sum, the self-interaction potentials considered here were shown to be
simple enough so as to obtain exact and physically sensible hairy black hole
solutions, and at the same time, sufficiently involved in order to provide
analytic solitons whose masses are fixed and determined by nontrivial
functions of the self-interaction parameters, as it can be explicitly seen
from eqs. (\ref{Msol I}), (\ref{mass lambda nu}). The link between soliton
masses and the entropy of their corresponding hairy black hole entropies is
discussed next.

\section{Microscopic entropy of hairy black holes: Soliton mass and its role
in the asymptotic growth of the number of states.}

\label{Section4}

The class of hairy black holes found here was shown to have positive mass and
it can be seen that they share many of the features with the one previously
found in \cite{HMTZ-2+1}. In fact, for any of the self interactions discussed
above, $V_{0,\nu}$ and $V_{\lambda,\nu}$, it occurs that for some fixed value
of the energy, the same theory admits the existence of at least two different
static and circularly symmetric black holes. Namely, for a precise value of
the mass, apart from the hairy black hole that is dressed with a nontrivial
scalar field, in vacuum one may also have the static BTZ black hole.
Furthermore, it is worth highlighting that, since both black holes depend on a
single integration constant, the hairy and BTZ black holes cannot be smoothly
deformed into each other, due to that fact that for a fixed mass, the scalar
field cannot be switched off. Following \cite{CMT}, this observation naturally
suggests that the hairy and the vacuum black holes belong to different
disconnected sectors. In the vacuum sector, the energy spectrum of the static
BTZ black hole possesses a continuous part bounded from below by zero, a gap
describing naked conical singularities, and a ground state that corresponds to
AdS spacetime, having a negative mass given by
\begin{equation}
M_{0}=\frac{2}{l}\tilde{\Delta}_{0}^{\pm}=-\frac{c^{\pm}}{12l}=-\frac{1}%
{8G}\ . \label{MAdS}%
\end{equation}
In the hairy sector the situation is similar, since the energy spectrum also
consists of a continuous part being bounded from below by zero that describes
the hairy black holes, a gap that corresponds to naked singularities, and
remarkably, a ground state that turns out to be consistently identified with
the solitons described above. Indeed, the solitons possess negative fixed
masses, given by eqs. (\ref{Msol I}) and (\ref{mass lambda nu}), being
completely determined by the fundamental constants of the theory. The soliton
solutions were also found to be smooth, regular everywhere, and devoid of
integration constants. Furthermore, they naturally provide the completion of
the hairy sector spectrum, since they not only fulfill the same asymptotic
conditions as the hairy black holes, but they also have precisely the same
boundary conditions, because the value of the constant $\alpha$, in eq.
(\ref{asympt scalar general}) coincides for both kinds of configurations.
Besides, unitarity of the dual theory for $c^{\pm}>1$ (see e.g. \cite{Libros}%
), together with the fact that the asymptotic growth of the number of states,
given by (\ref{Cardy super reloaded}), is well defined only for negative
lowest eigenvalues of the shifted Virasoro operators, impose the following
bounds on $\tilde{\Delta}_{0}^{\pm}$:%
\begin{equation}
-\frac{c^{\pm}}{24}\leq\tilde{\Delta}_{0}^{\pm}<0\text{\ .} \label{L0 bound}%
\end{equation}
Remarkably, full agreement is found from the bulk theory, since as expressed
by eq. (\ref{boundgroundstate}) this bound is precisely fulfilled by the
soliton mass that corresponds to the ground state of the hairy sector, and
according to (\ref{MAdS}) it is saturated in vacuum.

According to \cite{CMT}, the semiclassical entropy of the black holes under
consideration can then be suitably reproduced in terms of the microscopic
counting provided the ground state configuration is identified as the soliton
or the AdS spacetime, for the hairy and vacuum sectors, respectively.

Therefore, in the vacuum sector, since the lowest eigenvalues of the Virasoro
operators $\tilde{\Delta}_{0}^{\pm}$, are given by eq. (\ref{MAdS}), as
explained in introduction, the asymptotic growth of the number of states given
by (\ref{Cardy super reloaded}) reduces its standard form in eq.
(\ref{Cardy simple}). Thus, by virtue of (\ref{delta mas menos}) the
semiclassical entropy of the BTZ black hole is exactly reproduced as in Ref.
\cite{Strominger}.

The microscopic entropy of the hairy black holes discussed here can then be
obtained assuming that $\tilde{\Delta}_{0}^{\pm}$ are determined by the global
charges of their corresponding solitons, which according to eq.
(\ref{delta mas menos}) are given by $\tilde{\Delta}_{0}^{\pm}=\frac{l}%
{2}M_{0}$, where $M_{0}$ stands for the soliton mass. In the case of static
hairy black holes, as the ones discussed here, the asymptotic growth of the
number of states, given by eq. (\ref{Cardy super reloaded}), reduces to%
\begin{equation}
S=4\pi l\sqrt{-M_{0}M}\ , \label{entropy}%
\end{equation}
where $M$ is the hairy black hole mass. For the uniparametric potential
$V_{0,\nu}$, the semiclassical entropy of the hairy black hole, given by
(\ref{entropy uni}) is then successfully reproduced from a microscopic
counting once the black hole and soliton masses, given by (\ref{M1}) and
(\ref{Msol I}) are replaced into eq. (\ref{entropy}). Explicitly, this reads%
\[
S=4\pi l\sqrt{\frac{\Phi_{\nu}^{2}}{4G\nu}\times\frac{\nu a^{2}}{16Gl^{2}}%
}=\frac{\pi\Phi_{\nu}}{2G}a=\frac{A}{4G}\ .
\]
Analogously, for the more generic potential $V_{\lambda,\nu}$, taking into
account that the black hole and soliton masses are given by (\ref{mii}) and
(\ref{mass lambda nu}) respectively, formula (\ref{entropy}) reduces to%
\[
S=4\pi l\sqrt{\frac{\Upsilon_{\lambda,\nu}^{2}}{4G\nu}\times\frac{b^{2}%
}{16Gl^{2}}\frac{(\lambda-1)^{2}}{\lambda^{2}}\nu}=\frac{\pi}{2G}%
(1-\lambda)\frac{b}{\lambda}\Upsilon_{\lambda,\nu}=\frac{A}{4G}\ ,
\]
in full agreement with (\ref{Stwo}).

\section{Final remarks}

\label{discussion}

New asymptotically AdS hairy black holes and solitons were shown to exist for
General Relativity minimally coupled to a self-interacting scalar field in
three dimensions. Different self-interaction potentials were engineered in
order to obtain nontrivial analytic results, with the purpose of testing the
robustness of regarding the soliton as the ground state of the hairy sector,
and its key role in a microscopic counting of hairy black hole entropy. Our
results then confirm that this proposal successfully goes beyond the example
previously discussed in \cite{CMT} and naturally point towards the fact that
this mechanism should correspond to a generic feature of hairy black
holes\footnote{It is worth pointing out that similar results have also been
recently found in \cite{PTT, GTT} in the context of BHT\ massive gravity
\cite{BHT1, BHT2}; namely for asymptotically AdS hairy black holes and
solitons in vacuum \cite{OTT, GOTT}, as well as for black holes and solitons
with Lifshitz asymptotics \cite{ABGGH, GTT}.}.

In the microcanonical ensemble, i.e., for a fixed value of the mass, since the
theory admits hairy and vacuum black holes, it is natural to wonder which is
the preferred configuration. By virtue of eqs. (\ref{entropy}) and
(\ref{boundgroundstate}) (or equivalently (\ref{L0 bound})), the quotient of
the entropies of the vacuum and hairy black holes fulfills%
\[
\frac{S_{BTZ}}{S_{\textrm{hbh}}}=\sqrt{\frac{M_{AdS}}{M_{\textrm{sol}}}}>1\ ,
\]
where $M_{AdS}$, and $M_{\textrm{sol}}$ stand for the mass of AdS and the soliton,
respectively. Therefore, the vacuum black hole turns out to be the
thermodynamically preferred configuration. This result could be readily
extended for the (grand) canonical ensemble, as well as for the rotating case
through applying a boost in the \textquotedblleft$t-\varphi$\textquotedblright%
\ cylinder. It would then also be interesting to compare it with the one that
could be obtained from the mechanical stability of the hairy solutions.

As an ending remark, an interesting feature of the hairy black holes reported
here and their corresponding solitons that is worth to be highlighted, is that
their Euclidean continuations turn out to be diffeomorphic provided their
temperatures are related by an $S$-modular transformation.

\acknowledgments The authors thank Gaston Giribet, Marc Henneaux, Joaquim Gomis, Gabor
Kunstatter, Alfredo P\'{e}rez, David Tempo and Jorge Zanelli for useful discussions. F.
C. wishes to thank the kind hospitality at Universidad de Buenos Aires. F.C.
and C.M. thank the Conicyt grant 79112034 for financial support. This work has
been partially funded by the following Fondecyt grants: 1085322, 1095098,
1100755, 3100123, and by the Conicyt grant 
Anillo ACT-91: ``Southern Theoretical Physics Laboratory" (STPLab). The
Centro de Estudios Cient\'{\i}ficos (CECs) is funded by the Chilean government
through the Centers of Excellence Base Financing Program of Conicyt.

\appendix

\section{The functions $\Phi_{\nu}$ and $\Upsilon_{\lambda,\nu}$}
\label{apendiceA}

As shown in sections \ref{Section2} and \ref{Section3}, the functions
$\Phi_{\nu}$ and $\Upsilon_{\lambda,\nu}$ become relevant in order to
analytically describe the geometric and physical properties of the hairy black
holes and solitons that correspond to the self interactions $V_{0,\nu}$ and
$V_{\lambda,\nu}$, respectively. Some of their useful properties in this
context are detailed in this appendix.

\bigskip

{\Large $\cdot$} The function $\Phi_{\nu}$ is defined in terms of the Lambert
W function according to eq. (\ref{phinu}). It is a monotonically increasing
function of the parameter $\nu$, as it is depicted in Fig. 1. Its behaviour
around $\nu\rightarrow0$ is given by

\begin{figure}[h]
\centering
\includegraphics[scale=0.85]{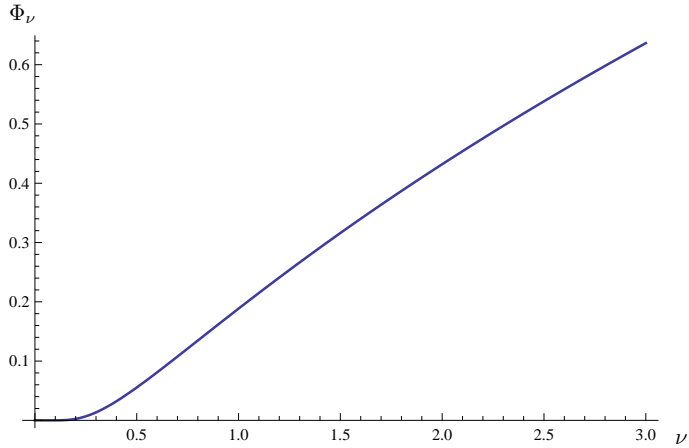}\caption{{Plot of the monotonically
increasing function $\Phi_{\nu}$.}}%
\end{figure}%

\begin{equation}
\Phi_{\nu}\xrightarrow[\nu\rightarrow{}0 ]\,e^{-1-\frac{1}{\nu}}%
+2e^{-2-\frac{2}{\nu}}+\frac{9}{2}e^{-3-\frac{3}{\nu}}+\cdots\ ,
\label{phicero}%
\end{equation}
while for $\nu\rightarrow\infty$, reads
\begin{equation}
\Phi_{\nu}\xrightarrow[\nu\rightarrow{}\infty]\,{\sqrt{\frac{\nu}{2}}-\frac
{2}{3}+\mathcal{O}\left(  \nu^{-1/2}\right)  \ }. \label{phiinfi}%
\end{equation}
Therefore, by virtue of eqs. (\ref{phicero}) and (\ref{phiinfi}), it can be
readily checked that the bound (\ref{Bound Phi}) is fulfilled.

\begin{figure}[h]
\centering
\includegraphics[scale=0.8]{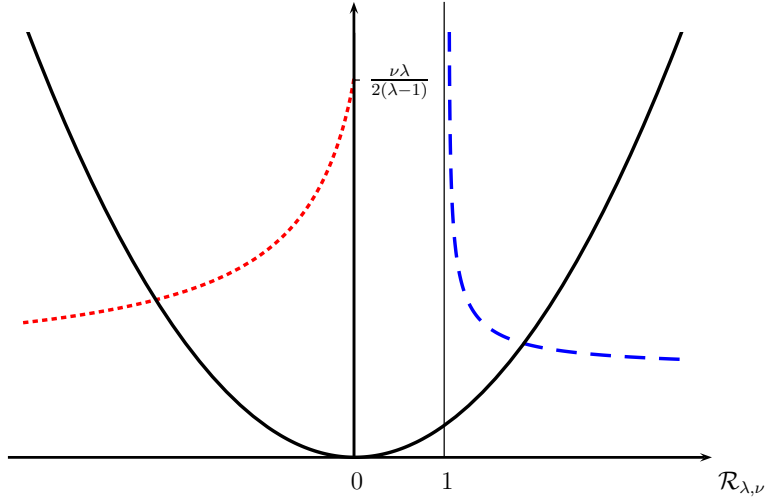}\caption{The plot represents the real
zeros of eq. (\ref{rdef}), which reads $\mathcal{R}_{\lambda,\nu}^{2}
=\frac{\nu}{(\lambda-1)}\left[  \frac{\lambda}{2}+\mathcal{R}_{\lambda,\nu
}+\mathcal{R}_{\lambda,\nu}^{2}\log\left(  1-\frac{1}{\mathcal{R}_{\lambda
,\nu}}\right)  \right]  $. The solid line corresponds to $\mathcal{R}_{\lambda,\nu}^{2}$, while the function at the right hand side, for the case
$\lambda>1$, is depicted by the dotted line, and for $\lambda<1$ is given by
the dashed line. The intersection of these curves then shows the existence of
a real root of (\ref{rdef}), which for $\lambda>1$ lies within the range
$-\infty<\mathcal{R}_{\lambda,\nu}<0$. This root approaches to $\sqrt
{\frac{\nu\lambda}{2(\lambda-1)}}$ for $\nu\rightarrow0$. For $\lambda<1$, the
plot shows that the real root is in the range $1<\mathcal{R}_{\lambda,\nu}<\infty$. }%
\end{figure}

\bigskip

{\Large $\cdot$} The function $\Upsilon_{\lambda,\nu}$ is defined in terms of
the function $\mathcal{R}_{\lambda,\nu}$ through eq. (\ref{Upsilon}), where
$\mathcal{R}_{\lambda,\nu}$ stands for the real root of (\ref{rdef}). Further
details about $\mathcal{R}_{\lambda,\nu}$ are revisited in Fig. 2. It is
simple to verify that it fulfills $\Upsilon_{\lambda,\nu}(1-\lambda)>0$. In
order to prove that the bound (\ref{bound upsilon}) holds, which reads%
\begin{equation}
\Upsilon_{\lambda,\nu}^{2}<\frac{\nu}{2}\ ,
\end{equation}
it is useful to recall eq. (\ref{rdef}), so that this inequality is
equivalently expressed as:
\begin{align}
F(\mathcal{R}_{\lambda,\nu})>0  &  \quad\mbox{if}\quad\lambda>1\ ,
\label{ine1}\\
F(\mathcal{R}_{\lambda,\nu})<0  &  \quad\mbox{if}\quad\lambda<1\ ,
\label{ine2}%
\end{align}
where%
\begin{equation}
F(\mathcal{R}_{\lambda,\nu}):=\frac{(\lambda+(1-\lambda)\mathcal{R}%
_{\lambda,\nu})^{2}}{2(\lambda-1)(\mathcal{R}_{\lambda,\nu}-1)^{2}}%
-\frac{\lambda}{2}-\mathcal{R}_{\lambda,\nu}-\mathcal{R}_{\lambda,\nu}^{2}%
\log\left(  1-\frac{1}{\mathcal{R}_{\lambda,\nu}}\right)  \ .
\end{equation}

In the case of $\lambda>1$, when $\mathcal{R}_{\lambda,\nu}\rightarrow-\infty$
the function $F(\mathcal{R}_{\lambda,\nu})$ approaches to $-\frac
{2}{3\mathcal{R}_{\lambda,\nu}}>0$, while for $\mathcal{R}_{\lambda,\nu
}\rightarrow0$ tends to $F(\mathcal{R}_{\lambda,\nu})\rightarrow\frac{\lambda
}{2(\lambda-1)}>0$. Hence, since the function $F(\mathcal{R}_{\lambda,\nu})$
is monotonically increasing in the range $-\infty<\mathcal{R}_{\lambda,\nu}%
<0$, the inequality (\ref{ine1}) holds.

Finally, for $\lambda<1$, when $\mathcal{R}_{\lambda,\nu}\rightarrow\infty$
the function $F(\mathcal{R}_{\lambda,\nu})$ tends to $-\frac{2}{3\mathcal{R}%
_{\lambda,\nu}}<0$, while if $\mathcal{R}_{\lambda,\nu}\rightarrow1$ it
approaches to $F(\mathcal{R}_{\lambda,\nu})\rightarrow\frac{1}{2(\lambda
-1)(\mathcal{R}_{\lambda,\nu}-1)^{2}}<0$. Therefore, as $F(\mathcal{R}%
_{\lambda,\nu})$ is a monotonically increasing function in the range
$1<\mathcal{R}_{\lambda,\nu}<\infty$, the inequality (\ref{ine2}) is satisfied.

\bigskip

It is simple to verify that in the case of $\lambda=0$, the function
$\Upsilon_{\lambda,\nu}$ reduces to $\Phi_{\nu}$ defined above, i.e.,%
\begin{equation}
\Upsilon_{0,\nu}=\Phi_{\nu}\,.
\end{equation}
Following the same procedure that allows to recover the potential
$V_{1,\tilde{\nu}}$ in (\ref{Potential}) from $V_{\lambda,\nu}$ in
(\ref{V lambda nu}), that consists on redefining $\nu=6\frac{\tilde{\nu}%
+1}{(\lambda-1)^{2}}$, and then taking the limit $\lambda\rightarrow1$, the function $\Upsilon_{\lambda,\nu}$ can be shown to fulfill
\begin{equation}
\lim_{\lambda\rightarrow1}\ (\lambda-1)^{2}\Upsilon_{\lambda,6(1+\tilde{\nu
})/(\lambda-1)^{2}}^{2}=\Theta_{\tilde{\nu}}^{2}\ ,
\end{equation}
where
\begin{equation}
\Theta_{\tilde{\nu}}:=2(z\bar{z})^{\frac{2}{3}}\frac{z^{\frac{2}{3}}-\bar
{z}^{\frac{2}{3}}}{z-\bar{z}},~\text{\textrm{with}}\quad z=1+i\sqrt{\tilde
{\nu}}.\label{Schuster}%
\end{equation}
As shown in \cite{HMTZ-2+1} and \cite{CMT}, the function $\Theta_{\tilde{\nu}%
}$ is the relevant one in order to obtain an analytic description of the hairy
black holes and solitons that correspond to the self interaction
$V_{1,\tilde{\nu}}$.

\section{Solutions in the conformal (Jordan) frame}
\label{apendiceB}

The three-dimensional hairy black hole and soliton solutions for a
self-interacting scalar field minimally coupled to General Relativity
discussed here, acquire an appealing form in the conformal frame. The action in eq. (\ref{Action}), after applying a conformal
transformation, followed by a scalar field redefinition of the form
\begin{equation}
\hat{g}_{\mu\nu}=\left(  1-\hat{\phi}^{2}\right)  ^{-2}g_{\mu\nu}%
\quad\mbox{and}\quad\hat{\phi}=\tanh\left(  \phi\right)  \ , \label{map}%
\end{equation}
reduces to the one for General Relativity with cosmological constant and a
conformally coupled self-interacting scalar field, given by
\begin{equation}
I[\hat{g},\hat{\phi}]=\frac{1}{\pi G}\int d^{3}x\sqrt{-\hat{g}}\left(
\frac{\hat{R}+2l^{-2}}{16}-\frac{1}{2}(\nabla\hat{\phi})^{2}-\frac{1}{16}%
\hat{R}\hat{\phi}^{2}-\hat{V}(\hat{\phi})\right)  \;. \label{accionconformal}%
\end{equation}
It can be shown that the potentials $\hat{V}_{0,\nu}$, $\hat{V}_{1,\nu}$, and
$\hat{V}_{\lambda,\nu}$ coming from their corresponding counterparts in the
Einstein frame discussed above, given by eqs. (\ref{V 0 nu}), (\ref{Potential}%
) and (\ref{V lambda nu}), respectively, are the only three branches of self
interactions that are compatible with static and spherically symmetric
solutions of the form $d\hat{s}^{2}=-f(\rho)dt^{2}+\frac{d\rho^{2}}{f(\rho
)}+\rho^{2}d\theta^{2}$, $\hat{\phi}=\hat{\phi}(\rho)$. As shown below, by means of 
(\ref{map}), this
class of solutions corresponds to the hairy black holes described in sections
\ref{Black hole I} and \ref{Black hole II} as well as the ones  in
Refs. \cite{Martinez:1996gn} and \cite{HMTZ-2+1} in the conformal frame. The
explicit form of the self interactions $\hat{V}_{0,\nu}$, $\hat{V}_{1,\nu}$,
and $\hat{V}_{\lambda,\nu}$, including also their corresponding soliton
solutions are discussed in what follows.

\subsection{Solutions for $\hat{V}_{1,\nu}$}

In the conformal frame, the self interaction (\ref{Potential}) is mapped into%
\begin{equation}
\hat{V}_{1,\nu}=-\frac{\nu}{8l^{2}}\hat{\phi}^{6}\ . \label{Vhat 1 nu}%
\end{equation}

This potential is the only one that turns out to be singled out by requiring
the matter piece of the action (\ref{accionconformal}) to be conformally
invariant; i.e., unchanged under local rescalings of the form $\hat{g}_{\mu
\nu}\rightarrow\lambda^{2}(x)\hat{g}_{\mu\nu}$, and $\hat{\phi}\rightarrow
\lambda^{-1/2}(x)\hat{\phi}$. In the case of $\nu\geq-1$, hairy black holes
solutions were found in \cite{HMTZ-2+1}, which reduces to the one previously
found in \cite{Martinez:1996gn} for $\nu=0$. Solutions describing solitons
were also found for the same range of the self-interaction coupling in
\cite{CMT}.

\subsection{Solutions for $\hat{V}_{0,\nu}$}

The self interaction (\ref{V 0 nu}) in the conformal frame, is mapped into%
\begin{equation}
\hat{V}_{0,\nu}=\frac{\nu}{16l^{2}}\frac{\hat{\phi}^{2}-2}{1-\hat{\phi}^{2}%
}\hat{\phi}^{2}-\frac{\nu}{8l^{2}}\ln\left(  1-\hat{\phi}^{2}\right)  \ ,
\label{V1}%
\end{equation}
whose behavior around $\hat{\phi}=0$ reads%
\begin{equation}
\hat{V}(\hat{\phi})\sim-\frac{\nu}{48l^{2}}\hat{\phi}^{6}-\frac{\nu}{32l^{2}%
}\hat{\phi}^{8}+\mathcal{O}(\hat{\phi}^{10})\ .
\end{equation}

\textbf{Hairy black hole:} In this case the metric is given by
\begin{equation}
d\hat{s}^{2}=-f(\rho)dt^{2}+\frac{d\rho^{2}}{f(\rho)}+\rho^{2}d\theta^{2}\ ,
\end{equation}
with
\begin{equation}
f(\rho)=\frac{\rho^{2}}{l^{2}}+\frac{\nu}{l^{2}}\left(  a\rho+\rho^{2}%
\ln\left(  1-\frac{a}{\rho}\right)  \right)  \ ,\label{simplest}%
\end{equation}
where the coordinates range as $-\infty<t<\infty$, $a<\rho<\infty$,
$0\leq\varphi<2\pi$. The scalar field also acquires a very simple form, that
reads
\begin{equation}
\hat{\phi}(\rho)=\sqrt{\frac{a}{\rho}}\ .
\end{equation}
The hairy black hole solution,  then depends on a single non-negative
integration constant $a$, that parametrizes the location of the event horizon, for $\nu>0$, 
at $\rho=\rho_{+}$, with
\begin{equation}
\rho_{+}=\frac{a}{1+W(-e^{-1-\frac{1}{\nu}})}\ ,
\end{equation}
where $W$ stands for the Lambert W function.

\textbf{Soliton:} The potential (\ref{V1}) also admits an additional exact
solution. The scalar field is given by%
\begin{equation}
\hat{\phi}(\rho)=\sqrt{\frac{1}{\rho}}\ ,
\end{equation}
and the metric reads%
\begin{equation}
d\hat{s}^{2}=-\rho^{2}dt^{2}+\frac{d\rho^{2}}{g(\rho)}+\frac{4l^{4}\Phi_{\nu
}^{2}}{\nu^{2}}g(\rho)d\varphi^{2}\ ,
\end{equation}
with%
\begin{equation}
g(\rho)=\frac{\rho^{2}}{l^{2}}+\frac{\nu}{l^{2}}\left(  \rho+\rho^{2}%
\ln\left(  1-\frac{1}{\rho}\right)  \right)
\end{equation}
Note that the solution is devoid of integration constants. The coordinates
range according to $1+\Phi_{\nu}\leq\rho<\infty$, $-\infty<t<\infty$, and
$0\leq\varphi<2\pi$. The metric and the scalar field are regular everywhere.
Since the mass does not depend on the choice of frame, it is given by
(\ref{Msol I}) and the solution then describes a soliton.

\subsection{Solutions for $\hat{V}_{\lambda,\nu}$}

In the conformal frame, the two-parametric potential (\ref{V lambda nu}) is
mapped into%
\begin{align}
\hat{V}_{\lambda,\nu}(\hat{\phi})  &  =\frac{\lambda^{2}}{8l^{2}}\hat{\phi
}^{6}-\frac{\nu}{8l^{2}(\lambda-1)}\left(  1-\lambda^{2}\phi^{6}\right)
\ln\left(  \frac{1-\lambda\phi^{2}}{1-\phi^{2}}\right) \nonumber\\
&  -\frac{\nu}{8l^{2}}\frac{1}{1-\phi^{2}}\left(  \phi^{2}+\frac{\lambda-1}%
{2}\phi^{4}+\frac{\lambda(\lambda-1)}{2}\phi^{6}-\frac{\lambda(\lambda+1)}%
{2}\phi^{8}\right)  \ , \label{V2}%
\end{align}
whose behavior around $\hat{\phi}=0$ is of the form
\begin{equation}
V_{\lambda,\nu}(\hat{\phi})=-\frac{\nu(\lambda-1)^{2}-6\lambda^{2}}{48l^{2}%
}\hat{\phi}^{6}+\frac{\nu(\lambda-1)^{3}}{32l^{2}}\hat{\phi}^{8}%
+\mathcal{O}(\hat{\phi}^{10})\ .\nonumber
\end{equation}

\textbf{Hairy black hole:} The field equations derived from
(\ref{accionconformal}) admit a static circularly symmetric solution whose
metric is given by
\begin{equation}
d\hat{s}^{2}=-f(r)dt^{2}+\frac{dr^{2}}{f(r)}+r^{2}d\varphi^{2}\ ,
\end{equation}
where $f(r)$ is expressed in eq. (\ref{fr}), and the scalar field reads%
\begin{equation}
\hat{\phi}(r)=\sqrt{\frac{b}{\lambda(r+b)}}\ .
\end{equation}
The solution then depends on a single integration constant $b$, and it is
well-defined provided $b/\lambda>0$. The coordinates range as $-\infty
<t<\infty$, $r_{s}<r<\infty$, and $0\leq\varphi<2\pi$. This solution describes
a hairy black hole for $\nu>0$, with an event horizon at $r=r_{+}$, with
$r_{+}$ given by (\ref{horizon2}), which surrounds the singularity at
$r=r_{s}$, located precisely as explained in Section \ref{Black hole II}.

\textbf{Soliton:} The self interaction (\ref{V2}) also admits soliton solution
described by the metric%
\begin{equation}
d\hat{s}^{2}=-\rho^{2}dt^{2}+\frac{d\rho^{2}}{g(\rho)}+l^{2}\lambda^{2}
\gamma_{\lambda,\nu}^{2}g(\rho)d\varphi^{2}\,,
\end{equation}
with%
\begin{equation}
g(\rho)=\frac{\rho^{2}}{l^{2}}-\frac{\nu}{\lambda(\lambda-1)l^{2}}\left(
\frac{(\lambda-1)^{2}}{2}-(\lambda-1)\rho+\lambda\rho^{2}\ln\left(
1+\frac{(\lambda-1)}{\lambda\rho}\right)  \right)  \,,
\end{equation}
where the scalar field is given by%
\begin{equation}
\hat{\phi}(\rho)=\sqrt{\frac{1}{\lambda(\rho+1)}}\ .
\end{equation}
The soliton possesses no integration constants, and taking in account that the
coordinates range as $\frac{1}{\lambda}+\frac{(1-\lambda)\Upsilon_{\lambda
,\nu}}{\lambda}\leq\rho<\infty$, $-\infty<t<\infty$, $0\leq\varphi<2\pi$, it
is simple to verify that the solution is is regular everywhere.

\end{document}